\documentclass[12pt]{article}
\usepackage{eurosym}
\usepackage{amssymb,amsmath}
\usepackage[dvips]{color}
\usepackage{amscd}
\usepackage{tikz}

\usetikzlibrary{patterns}

\def\map{\longrightarrow}
\def\ch{\mathrm{ch}}
\def\and{\mathrm{and}}

\def\sO{\mathcal O}

\def\bP{\mathbb P}
\def\bZ{\mathbb Z}

\def\tX{\tilde {X}}

\def\map{\longrightarrow}

\newtheorem{prop}{Proposition}

\newtheorem{thm}{Theorem}

\newcommand{\be}{\begin{equation}}
\newcommand{\ee}{\end{equation}}
\newcommand{\bea}{\begin{eqnarray}}
\newcommand{\eea}{\end{eqnarray}}
\newcommand{\beas}{\begin{eqnarray*}}
\newcommand{\eeas}{\end{eqnarray*}}
\newcommand{\ba}{\begin{array}}
\newcommand{\ea}{\end{array}}

\newcommand{\nbox}{{\,\lower0.9pt\vbox{\hrule \hbox{\vrule height 0.2 cm \hskip 0.19 cm \vrule height 0.2 cm}\hrule}\,}}

\def\href#1#2{#2}
\input{tcilatex}

\begin{document}

\begin{titlepage}
\hfill
\vbox{
    \halign{#\hfil         \cr
           } 
      }  

\hbox to \hsize{{}\hss \vtop{ \hbox{}

}}

%

\vspace*{20mm}
\begin{center}

{\large \textbf{Heterotic String Compactification and New Vector Bundles} }

{\Large \vspace{ 20mm} }

{\normalsize {Hai Lin${}^{1,2,3}$, Baosen Wu${}^{1}$, and Shing-Tung Yau${}^{1,2,3}$}  }

{\normalsize \vspace{10mm} }

{\small \emph{${}^1$\textit{Department of
Mathematics, Harvard University, Cambridge, MA 02138, USA
}} }

{\normalsize \vspace{0.2cm} }

{\small \emph{${}^2$\textit{Center of Mathematical Sciences and Applications, Harvard University, \\Cambridge, MA 02138, USA
}} }

{\normalsize \vspace{0.2cm} }

{\small \emph{$^3$\textit{Department of
Physics, Harvard University, Cambridge, MA 02138, USA
\\
}} }

{\normalsize \vspace{0.4cm} }

%
\end{center}

\begin{abstract}

We propose a construction of K\"{a}hler and non-K\"{a}hler Calabi-Yau manifolds by branched double covers of twistor spaces. In this construction we use the twistor spaces of four-manifolds with self-dual conformal structures, with the examples of connected sum of $n$ $\mathbb{P}^{2}$s. We also construct $K3$-fibered Calabi-Yau manifolds from the branched double covers of the blow-ups of the twistor spaces. These manifolds can be used in heterotic string compactifications to four dimensions. We also construct stable and polystable vector bundles. Some classes of these vector bundles can give rise to supersymmetric grand unified models with three generations of quarks and leptons in four dimensions.

\end{abstract}

\end{titlepage}

\vskip 1cm

\section{Introduction}

\vspace{1pt}\renewcommand{\theequation}{1.\arabic{equation}} %
\setcounter{equation}{0}

Superstring compactifications from heterotic string theory to four
dimensions give a promising approach to find realistic Standard Model like
particle physics with three generations of quarks and leptons. In the
heterotic string compactification, the higher dimensional spacetime is a
product of the Minkowski four-manifold and the internal six dimensional
manifold, see for example \cite{Candelas:1985en, Witten:1985bz, Strominger:1986uh}. One of the standard approaches has been the
compactification of the heterotic string on smooth Calabi-Yau three-folds with
holomorphic vector bundles. These bundles break the $E_{8}$ gauge
theory down to $E_{6}$, $SO(10)$ and $SU(5)$ grand unified theories. Many classes of models
using general holomorphic vector bundles on the internal manifold can lead to
any of the above three grand unified groups. These unified gauge groups can further break down to the Standard Model gauge
group, for example with Wilson line turned on as a usual method. In the context of the heterotic string
compactifications, progresses of building phenomenologically viable models have been made in for example \cite{Donagi:2000zs, Anderson:2012yf, Anderson:2013xka, Braun:2009qy, Cleaver:2011ir, Lebedev:2007hv, Gabella:2008id, Bouchard:2006dn, Anderson:2011ns, Blumenhagen:2006ux, Weigand:2006yj} and references therein.

The heterotic string theory contains the vector bundle degree of freedom,
and the first and second Chern classes of the vector bundle on the
Calabi-Yau manifolds need to satisfy nontrivial constraints. In the usual
Calabi-Yau compacitifcations, the first Chern class of the vector bundle
vanishes due to the Hermitian Yang-Mills equations, and the second Chern
class of the vector bundle equals the second Chern class of the tangent
bundle, up to a total effective curve class from fivebranes.

Heterotic string compactification on a simply connected Calabi-Yau manifold
can also been considered \cite{Blumenhagen:2006ux, Andreas:1998ei,
Curio:1998vu} with the examples of elliptic Calabi-Yau manifolds. The vector
bundles on elliptically fibered Calabi-Yau can be constructed by means of
spectral cover construction, see for example \cite{Friedman:1997ih,
Donagi:1997dp}. In some cases, freely acting involutions in elliptically
fibered Calabi-Yau threefolds with two sections, were also proposed, see for
example \cite{Andreas:2011cz, Andreas:1999ty, Curio:2004pf} and references
therein.

In this paper we construct K\"{a}hler and non-K\"{a}hler Calabi-Yau
manifolds which can be used in the heterotic string compactifications. We
construct them from branched double covers of twistor spaces of four-manifolds
with self-dual conformal structures. These twistor spaces have balanced
metrics. The manifolds as the branched covers solve the conformally balanced
equation. We also construct stable and polystable vector bundles on these
manifolds, which satisfy the anomaly cancellation condition and the
Hermitian-Yang-Mills equations. Among the vector bundles we construct, there
are those which gives supersymmetric grand unified models with three generations of quarks
and leptons.

The non-K\"{a}hler Calabi-Yau spaces here are complex three-folds with
trivial canonical bundle. They may play an important role in
the compactification of heterotic string theory to four dimensions. The
construction of these non-K\"{a}hler Calabi-Yau manifolds opens up more possibilities in
finding the vacuum corresponding to the Standard Model from the superstring
theory.

Our approach include both K\"{a}hler Calabi-Yau spaces and non-K\"{a}hler
Calabi-Yau spaces. More specifically, we constructed the K\"{a}hler and
non-K\"{a}hler Calabi-Yau spaces by branched double covers of twistor spaces, or
branched double covers of the blow-ups of the twsitor spaces. We consider the twistor spaces of four-manifolds $M^{4}$, for example,
the connected sum of $n$ $\mathbb{P}^{2}$s, which have self-dual conformal
structures. The branch locus of the double cover of the twistor space \textrm{Tw}$(M^{4})$ can be either smooth or singular. In the smooth case, the
branch locus is a divisor whose divisor class is twice of the divisor class
of $K3$ surface. In the singular case, the branch locus is a union of two $K3$
surfaces, and after blowing up along the singular locus, the resulting manifolds are Calabi-Yau manifolds
and are also $K3$ fibrations over $\mathbb{P}^{1}$.

We consider the exploration of a specific Standard Model gauge group or GUT
gauge group with three generations of quarks and leptons. To obtain $SU(5)$
and $SO(10)$ GUT groups as the subgroups of the $E_{8}$ group, we need to
construct rank 5 and rank 4 vector bundles, respectively. These vector
bundles are stable or polystable. The polystability or stability of the
vector bundle guarantees the existence of the solution to the
Hermitian-Yang-Mills equations, see for example \cite{LY
hermitian, Andreas:2010cv}.

We also propose a new approach to construct stable and polystable bundles. We
constructed rank 5 bundles with nonzero $c_{3}$, which include
the rank 5 bundles with $c_{3}=6$, corresponding to supersymmetric $SU(5)$ GUT
models with three generations of chiral fermions. After a GUT symmetry
breaking, some of the models can give rise to models with Standard Model
gauge groups and three generations of chiral fermions. These examples have
relevance to model buildings for obtaining phenomenologically viable
four-dimensional theory. This is also an example where the anomaly
cancellation condition is satisfied without adding fivebranes or effective
curve class. In another example, we constructed rank 4 bundles with
zero $c_{3}$, in which fivebranes or effective curve class are introduced in
the anomaly cancellation condition.

The organization of this paper is as follows. In Section \ref{sec_physical
constraints}, we describe the physical constraints of the relevant manifolds
and the Chern classes of the vector bundles. In Section \ref{sec_construction general aspects}, we propose a general procedure for
constructing the K\"{a}hler and non-K\"{a}hler Calabi-Yau manifolds by using
branched double covers of twistor spaces of four-manifolds, with the examples of the connected sum of $n$ $\mathbb{P}^{2}$s. Afterwards in Section \ref{sec_construction Kahler CY}, we construct
$K3$-fibered Calabi-Yau manifolds from the branched double covers of the blow ups of the
twistor spaces, and in particular the K\"{a}hler Calabi-Yau manifolds for the K\"{a}hler twistor spaces with the example of ${S^4}=0\mathbb{P}^{2}$.
Then in Section \ref{sec_construction non-Kahker}, we construct the case of
non-K\"{a}hler Calabi-Yau manifolds for the non-K\"{a}hler twistor spaces with the example of $2\mathbb{%
P}^{2}$. Finally we briefly discuss our results and make some
conclusions in Section \ref{sec_discussion}.

\label{sec_introduction}\vspace{1pt}

\section{Physical constraints}

\vspace{1pt}\renewcommand{\theequation}{2.\arabic{equation}} %
\setcounter{equation}{0}

\label{sec_physical constraints}\vspace{1pt}

The heterotic superstring theory can be compactified on a warped product of
Minkowski four-manifold and an internal complex three-fold. The complex
three-fold is endowed with a holomorphic $(3,0)~$three-form $\Omega $ and a
Hermitian $(1,1)$ form $\omega ~$associated with the Hermitian metric. The
background contains a vector bundle $V$, and the gauge fields $F$ of the
vector bundle satisfy the Hermitian Yang-Mills equations. The
background also satisfy the conformally balanced condition and the anomaly
cancellation condition. These above three conditions are described by
\begin{equation}
d(\left\Vert \Omega \right\Vert ~\omega ^{2})=0,\text{ \ \ }
\label{conformally balanced}
\end{equation}%
\begin{equation}
F^{1,1}\wedge \omega ^{2}=0\ \ \mathrm{and}~~F^{(2,0)}=F^{(0,2)}=0,
\label{Hermitian YM}
\end{equation}%
\begin{equation}
dH=\frac{\alpha ^{\prime }}{4}\left( \text{tr}(R\wedge R)-\text{tr}(
F\wedge F) \right) -[W].  \label{c2 equation}
\end{equation}%
The anomaly cancellation condition (\ref{c2 equation}) can also be
characterized as a modified Bianchi identity for the $H$-flux, in which the $%
[W]$ term corresponds to the fivebrane source term to the $H$-flux. These
equations are analyzed in details by for example \cite{Strominger:1986uh,
Fu:2006vj, Becker:2006et, Fu:2008ga, Becker:2009df}. The norm $\left\Vert
\Omega \right\Vert $ is defined by\ $\Omega \wedge \overline{\Omega }=-i%
\frac{4}{3}\left\Vert \Omega \right\Vert ^{2}$ $\omega ^{3}.~$The physical
fields are related to the holomorphic three-form and the Hermitian form of
the manifolds by
\begin{equation}
H=i(\overline{\partial }-\partial )\omega \text{, \ \ }e^{-2\phi
}=\left\Vert \Omega \right\Vert
\end{equation}%
and $dH=2i\partial \overline{\partial }\omega$. These complex three-folds
also preserve $N=1$ supersymmetry in four dimensions.

To have $SU(5)$ and $SO(10)$ grand unified groups as the subgroups of the $%
E_{8}$ group, one need to construct rank 5 and rank 4 vector bundles
respectively. The commutant group of the rank 5 bundle in the $E_{8}~$group
is the $SU(5)$ grand unified group, while the commutant group of the rank 4
bundle in the $E_{8}$ group is the $SO(10)$ grand unified group. The two
equations (\ref{Hermitian YM}) in the Hermitian Yang-Mills mean that the
vector bundle $V$ is holomorphic and $c_{1}(V)=0$. These equations have solutions when the vector bundles are stable or polystable \cite{Donaldson, UY}. The stability or polystability of the vector bundle guarantees the existence of the solution to the Hermitian-Yang-Mills equations, see for example \cite{LY
hermitian, Andreas:2010cv}.

Since our branched double cover construction produces Calabi-Yau manifolds
that are either K\"{a}hler or non-K\"{a}hler, the first Chern class of the
tangent bundle of $M$ is zero, that is $c_{1}(M)=0$. In both these cases in 
this paper, these manifolds are complex three-folds.

The heterotic string compactification with three generations of chiral
fermions have three physical constraints on the Chern classes of the vector
bundle $V$ as
\begin{equation}
c_{1}(V)=0,  \label{c1}
\end{equation}
\begin{equation}
c_{2}(V)=c_{2}(M)-[W],  \label{c2}
\end{equation}
\begin{equation}
c_{3}(V)=6.  \label{c3}
\end{equation}
The first two conditions (\ref{c1}) and (\ref{c2}) are necessary conditions
for consistent solutions. The $[W]$ is a total effective curve class coming
from fiverbanes. The third condition (\ref{c3}) is not a necessary condition
for general solutions, but is a condition for having three generations of
chiral fermions when reducing the model to four dimensions. The
configurations that satisfy all other constraints except the one for $%
c_{3}(V)$ are solutions to the heterotic string theory, though not having
three generations of chiral fermions. If the manifold $M$ has a
freely acting involution $\gamma ,$ of order $|\gamma|$, then by the index
theorem and the Riemann-Roch formula, the number of generations of the
chiral fermions are
\begin{equation}
\text{\ }N=\int \mathrm{ch}(V)\mathrm{Td}(M)=\frac{c_{3}(V)}{2|\gamma |}=3.
\end{equation}%
Here, the\ $\mathrm{ch}(V)$ is the Chern character and $\mathrm{Td}(M)$ is
the Todd class. Two relevant cases in our approach are $c_{3}(V)=6,|\gamma
|=1 $ and $c_{3}(V)=12,|\gamma |=2$.

Two common grand unified groups are $SU(5)$ and $SO(10)$. We will focus on
the construction of rank 5 bundles $V$ corresponding to supersymmetric $SU(5)$ GUT models.
In this case we have that
\begin{equation}
\wedge ^{5}V\cong \mathcal{O}_{M}.
\end{equation}%
The Higgs particles in these models can be in the representations \textbf{5}$%
_{H}$ or \textbf{\={5}}$_{H}$. The matter fields can be in the the
representations \textbf{\={5}} or\textbf{~10}. The representations \textbf{%
\={5}},\textbf{~5 }of the $SU(5)$ GUT model correspond to $H^{1}(M,\wedge
^{2}V)$ and $H^{1}(M,\wedge ^{2}V^{\vee })$ respectively, while the
representation \textbf{10} corresponds to $H^{1}(M,V)$. There exist
several types of couplings between the Higgs particles and the matter
particles, the \textbf{\={5}} \textbf{\={5} 10}, which corresponds to
the nonzero pairing
\begin{equation}
H^{1}(M,\wedge ^{2}V)\otimes H^{1}(M,\wedge ^{2}V)\otimes
H^{1}(M,V)\longrightarrow \mathbb{C,}
\end{equation}%
and the \textbf{10 10} \textbf{5}, which corresponds to the nonzero
pairing
\begin{equation}
H^{1}(M,V)\otimes H^{1}(M,V)\otimes H^{1}(M,\wedge ^{2}V^{\vee
})\longrightarrow \mathbb{C,}
\end{equation}%
where we have used $H^{3}(M,\wedge ^{5}V)\cong H^{3}(M,\mathcal{O}_{M})\cong
\mathbb{C}$. In addition, higher dimensional representations of the Higgs
particles in $SU(5)$ grand unified models are possible.

There are similar conditions for the rank 4 vector bundles $V$, corresponding to the supersymmetric $SO(10)$ GUT models. These have been discussed in detail in \cite{Thomas:1999iz}. In this case
\begin{equation}
\wedge^{4}V\cong \mathcal{O}_{M}.
\end{equation}%
The Higgs particles in these models can be in the representations \textbf{10}$_{H}$. The matter fields can be in the the representations \textbf{16}. The
representation \textbf{10} of the $SO(10)$ GUT model corresponds to
$H^{1}(M,\wedge^{2}V)$, and the representation \textbf{16} corresponds to $H^{1}(M,V)$. There exist couplings between the Higgs particles and the
matter particles, the \textbf{10} \textbf{16 16}, which corresponds to
the nonzero pairing
\begin{equation}
H^{1}(M,\wedge^{2}V)\otimes H^{1}(M,V)\otimes H^{1}(M,V)\longrightarrow
\mathbb{C,}
\end{equation}%
where we have used $H^{3}(M,\wedge^{4}V)\cong H^{3}(M,\mathcal{O}_{M})\cong
\mathbb{C}$. Again, higher dimensional representations of the Higgs
particles in $SO(10)$ grand unified models are possible.

\section{Construction of Calabi-Yau threefolds from twistor spaces}

\vspace{1pt}\renewcommand{\theequation}{3.\arabic{equation}} %
\setcounter{equation}{0}

\label{sec_construction general aspects}\vspace{1pt}

\vspace{1pt}

There are many ways to construct Calabi-Yau manifolds. Here we shall describe general ideas of constructing Calabi-Yau threefolds as double cover of twistor spaces. The resulting Calabi-Yau is often non-K\"{a}hler. The advantage of working on twistor spaces is that we can find natural balanced metrics on such Calabi-Yau threefolds. This section is devoted to general aspects of this approach. In the next two sections we shall describe specific examples and application to heterotic superstring theory.

\vspace{1pt}\vspace{1pt}

\subsection{Twistor spaces of connected sum of $\bP^{2}$s}

\vspace{1pt}\vspace{1pt}

\def\tw{\mathrm{Tw}}
\def\bF{\mathbb F}

We start by recalling some basic notions of twistor spaces of self-dual four-manifolds. For any oriented four-manifold $M^4$, its twistor space is defined as
\begin{equation}
\tw(M^4)=P\times_{SO(4)}SO(4)/U(2),
\end{equation}
where $P$ is the $SO(4)$ principal bundle of $M^4$. It was proved \cite{AHS78} that $M^4$ admits a self-dual conformal structure, that is, $W_-=0$ for Weyl tensor $W$, if and only if the natural almost complex structure on $\tw(M^4)$ is integrable. Taubes \cite{Taubes} showed that for any compact oriented four manifold, after taking connected sum with sufficiently many $\bP^2$s, the resulting four manifold admits a self-dual conformal structure.

In the sequel we assume $M^4$ is self-dual.  The twistor space has a natural differentiable map $\tw(M^4)\to M^4$ which is an $S^2$-fibration. Each fiber is a holomorphic $\bP^1$ with the induced holomorphic structure. In addition, there is a real structure, an anti-holomorphic map $\tau:\tw(M^4)\to \tw(M^4)$, preserving the fibration and induces the antipodal map on each fiber $S^2$.

Next we shall focus on four-manifolds $M^4=n\bP^2$, the connected sum of $n$ copies of $\bP^2$s. Floer \cite{Flo91} proved the existence of self-dual metrics on $n\bP^{2}$ by perturbation arguments. Shortly after that, Donaldson and Friedman \cite{DF89} gave an algebraic proof of a more general result by constructing its twistor space using deformation theory which will be sketched later.

The simplest example is $n=0$, that is, the $4$-sphere $S^4$. It is well-known that $\tw(S^4)=\bP^3$. Hitchin \cite{Hitchin} showed that in the compact case, the twistor space is K\"{a}hler if and only if it is $\bP^3$ or the complete flag manifold $\bF={\bP}(T_{\bP^2})$,
which are twistor spaces of standard conformal structures on $S^4$ and $\bP^2$ respectively.

For $n=2$ and $3$, Poon \cite{Poon86} analyzed in details the structure of the twistor spaces $\tw(n\bP^2)$. They are non-K\"{a}hler but turn out to be Moishezon, and there is a moduli of such self-dual conformal structures.

When $n\ge 4$, it becomes complicated. The twistor space can have algebraic dimension zero when $n$ is large. However, LeBrun \cite{Lebrun} gave an explicit conformal structure for all $n\bP^2$. In addition, the twistor spaces of these conformal structures are Moishezon and can be described explicitly. For other related work on twistor spaces of $n\bP^2$, see for example \cite{P-P, Lebrun Poon, Ped Poon, Lebrun86, Poon}.

\subsection{Donaldson-Friedman construction}

For later use, we recall briefly Donaldson-Friedman's construction of self-dual four-manifolds from
deformation of singular spaces. We shall focus on the case of $M^4=2\bP^2$.

Recall that the twistor space of $\bP^2$ with its Fubini-Study metric is the flag manifold $\bF={\bP}(T_{\bP^2})$. Let $\pi:\tilde{\bF}\to \bF$ be the blowup of $\bF$ along a real twistor line. Then the exceptional divisor is isomorphic to $\bP^1\times \bP^1$ with normal line bundle $\sO(1,-1)$.

Take two identical such flag manifolds $\bF$, labeled by $\bF_1$ and $\bF_2$. After blowing-up we obtain $\tilde{\bF}_1$ and $\tilde{\bF}_2$ with exceptional divisors $E_1$ and $E_2$, both isomorphic to $\bP^1\times \bP^1$. Note that $\bP^1\times \bP^1$ has an automorphism $u$ switching the two factors. We glue $\tilde{\bF}_1$ and $\tilde{\bF}_2$ along $E_1$ and $E_2$ via such automorphism $u$, and obtain
\begin{equation}
Z_0=\tilde{\bF}_1\cup_{\bP^1\times \bP^1} \tilde{\bF}_2
\end{equation}
which is simple normal crossing with singularity along $D=\bP^1\times \bP^1$. Furthermore, $Z_0$ admits a natural real structure.

On each component $\tilde{\bF}_i$ of $Z_0$, we denote the normal bundle of $D$ by $N_i$ for $i=1,2$. Then $N_1=\sO(1,-1)$ and $N_2=\sO(-1,1)$. Therefore $N_1\otimes N_2=\sO_D$; in other words, $Z_0$ is $d$-semistable.

Now we consider the general theory of global smoothing, and suppose we have a $d$-semistable space $Y=Y_{1}\cup_{D}Y_{2}$. Let $N_{i}$ be normal bundles of $D$ in $Y_{i}$. If the following spaces
\begin{enumerate}
\item $H^{1}(\sO_{D}),\,H^{2}(T_{Y_{i}}),$
\item $H^{p}(N_{i})\quad\text{\textrm{for all }}p,$
\item $H^{p}(T_{D}),\,p=1,2$
\end{enumerate}
are vanishing, then there is a global smoothing of $Y$ \cite{DF89}. Applying this criterion to the case $Z_0=\tilde{\bF}_1\cup_{\bP^1\times \bP^1} \tilde{\bF}_2$, we can verify that $Z_0$ admits a smoothing. In addition, there is a smoothing which preserves the real structure so that such a smoothing gives a twistor space $Z$. The corresponding four-manifold is then a self-dual structure on $2\bP^{2}$.

\subsection{Calabi-Yau as double cover of twistor spaces}

The double cover construction of Calabi-Yau works for any Fano threefold, or more generally, any smooth threefold $X$ so that $|-2K_X|$ admits a smooth divisor.

Given such a threefold $X$, a nontrivial section $s\in H^0(-2K_{X})$ defines a homomorphism $m:K_{X}\otimes
K_{X}\to \sO_{X}$, which in turn induces an algebraic structure on the sheaf $\sO_{X}\oplus K_{X}$. Denote
\begin{equation}
M = \mathbf{Spec}(\sO_{X}\oplus K_{X}).
\end{equation}
The natural map $f:M\to X$ is a double cover of $X$. Now suppose the section $s\in H^0(-2K_{X})$ defines a smooth divisor $B=s^{-1}(0)$, then $M$ is smooth and it is straightforward to show that the canonical bundle $K_{M}$ is trivial using adjunction formula. Moreover, one can verify that $B\subset X$ is the branch divisor of $f: M\to X$.

We shall also work on the case $B$ is not smooth. Suppose $B_i$ are smooth divisors linearly equivalent to $-K_X$, and $B=B_1\cup B_2$ with simple normal crossing singularity. Let $C=B_1\cap B_2$. Then we can still construct a double cover $f:M'\to X$ as before. However, $M'$ is not smooth. It has ordinary double point singularity along $f^{-1}(C)$. It is easy to verify that the dualizing sheaf of $M'$ is trivial, and a small resolution $M\to M'$ gives a smooth threefold $M$ with trivial canonical bundle.

The following is an equivalent point of view for this singular case. Let $\tilde X\to X$ be the blowup of $X$ along $C=B_1\cap B_2$. Then the anti-canonical divisor of $\tilde X$ is base point free and defines a fibration $\tilde X\to \bP^1$. Let $\tilde B_i$ be the proper transform of $B_i$. Then $\tilde B=\tilde B_1\cup \tilde B_2$ is a disjoint union and is the branch locus that we use to define a double cover of $\tilde X$. It turns out in this way we obtain the same manifold $M$ by the previous construction.

The singular case has an extra nice structure. Because $B_i$ is anti-canonical divisor, adjunction formula implies that it has trivial canonical bundle. If furthermore it is simply connected, then it is a $K3$ surface. Therefore we get a $K3$ fibration structure on the resulting double cover manifold.

We now focus on the case when $X$ are the spaces $\tw(n\bP^2)$ with LeBrun's conformal structure on $n\bP^2$. Explicit description of $X$ shows that $-2K_X$ has a section defines a smooth divisor, or simple normal crossing divisor of the type discussed above. One therefore obtains Calabi-Yau threefolds from double cover construction. These Calabi-Yau manifolds are Moishezon. For small $n$, particularly less than $4$, the projective model for $X$ is well known. In the next two sections of the paper, we shall work out more geometric structures for the case $n=0$ and $n=2$.

\subsection{Hermitian forms and conformally balanced metrics}


We denote $Z=\tw(n\bP^2)$ for LeBrun's conformal structure on $n\bP^2$. Then $n\bP^2$ has positive scalar curvature. There is a natural family of balanced Hermitian metrics on $Z$ with associated positive $(1,1)$-forms ${\omega_Z}$.

Let $f:M\to Z$ be a double cover branched along a smooth divisor $B$. We construct a balanced metric on the resulting double cover $M$ as follows. Recall that the map $\omega\to\omega^2$ defines a bijection between the cone of positive $(1,1)$-forms and positive $(2,2)$-forms \cite{Mic82}. It suffices to find a closed positive $(2,2)$-form on $M$.

Note that the pull-back $(2,2)$-form $f^*(\omega_Z^2)$ is positive away from the ramification divisor $f^{-1}(B)$, and it is closed. We modify it to a positive closed form in the following way. Let $L$ be the line bundle over $M$ with a section $s\in H^0(L)$ so that its zero locus is $f^{-1}(B)$. Since $f^{-1}(B)$ is projective and $L|_{f^{-1}(B)}$ is positive, one can find an Hermitian metric $h$ on $L$ so that the Chern form $c_{1}(L,h)$ is positive on a neighbourhood $U$ of the ramification divisor $f^{-1}(B)$. It follows that
\begin{equation}
c_{1}({L},h)^{2}|_{U}>0.
\end{equation}
On the other hand, $c_{1}({L},h)^{2}$ is bounded and
\begin{equation}
f^*(\omega_Z^2)|_{M\backslash U}>0.
\end{equation}
We can find a sufficiently large constant $C>0$ so that
\begin{equation}
\omega_M^{2}:=C\cdot f^*(\omega_Z^2)+c_{1}({L},h)^{2}
\end{equation}
is a positive closed $(2,2)$-form.

Having a balanced metric $\omega_M$ on $M$, we can reduce the conformally balanced equation $d({\left\Vert \Omega \right\Vert}_{\omega}\omega ^{2})=0$ to $\omega_M^2= {\left\Vert \Omega \right\Vert}_{\omega}\omega^2 $. This is essentially a complex Monge-Amp\`{e}re equation which is solvable by the method in \cite{Yau}.


\subsection{Deformation construction}

We describe a more general construction of Calabi-Yau threefolds by double cover of singular space and smoothing.

Again we let $\bF$ be the twistor space of $\bP^2$. Let $\pi :\tilde{\bF}\to \bF$ be the blow up of $\bF$ along a real twistor line $\ell$ with exceptional divisor $E$. Then $K_{\tilde{\bF}}=\pi^{\ast }K_{\bF}+E$. Hence the log canonical divisor of $\tilde \bF$ is $K_{\tilde{\bF}}+E=\pi^{\ast }K_{\bF}+2E$.

Recall that we defined
\begin{equation}
Z_0=\tilde{\bF}_1\cup_{\bP^1\times \bP^1} \tilde{\bF}_2
\end{equation}
with $D=E_1\cong E_2$ via an isomorphism $u$. The dualizing sheaf of $Z_0$ is a trivial line bundle coming from gluing of $K_{\tilde{\bF}_{i}}+E_i$ along their restrictions to $D$.

We construct double cover of the singular space $Z_0$ which is again a threefold with simple normal crossing singularity. A global smoothing of it gives a Calabi-Yau threefold.

More precisely, we start with double cover $Y\to \tilde{\bF}$ branched along $-2(K_{\tilde{\bF}}+E)$. Note that $\ell $ is a twistor line, we have $K_{\bF}\cdot \ell =-4$. Therefore, the restriction of $K_{\tilde{\bF}}+E$ to $E$ is isomorphic to $\mathcal{O}(-2,-2)=K_{E}$.

Let $S=Y\times_{\tilde{\bF}} E$. Then the projection $S\to E$ is a double cover branched over a divisor $-2K_E$. Hence $S$ is a $K3$ surface. In this way we obtain a pair $S\subset Y$ lying above $E\subset \tilde{\bF}$. The
normal bundle of $S$ in $Y$ is the pull back of $\sO(1,-1)$.

The following is the main result of this subsection:
\begin{prop}
Let $\pi:\tilde{\bF}\to \bF$ be the blowup of $\bF$ along a real twistor line $\ell$ with
exceptional divisor $E=\bP^1\times\bP^1$. Let $\tilde{\bF}_1$ and $\tilde{\bF}_2$ be two copies of $\tilde{\bF}$. Suppose we can find smooth divisors $D_i\subset \tilde{\bF}_i$ in the class $-2K_{\tilde{\bF}_i}-2E_i$
so that the intersection $D_i\cap E_i$ is smooth and invariant under the automorphism of $E_i$ switching the two factors. Then we obtain double covers $Y_i$ of $\tilde{\bF}_i$ and singular space $M_0=Y_1\cup Y_2$ gluing along a
$K3$, so that there exists a smoothing of $M_0$ to an (often non-K\"{a}hler) Calabi-Yau threefold.
\end{prop}


\subsection{Noncompact case}

One can also consider the noncompact Calabi-Yau manifolds constructed from the twistor spaces.
In \cite{Heckman:2013sfa}, a different double cover was taken, in which the
branch locus is a $K3$, and the resulting double cover is a positive curvature
manifold. In \cite{Heckman:2013sfa} then a noncompact Calabi-Yau can be
produced by deleting a divisor from this positive curvature manifold.

The double cover in this paper is different from the double cover in \cite{Heckman:2013sfa}
because the branch locus is of different divisor class. In
this paper, the twistor space is branched over a divisor that is in twice
the divisor class of $K3$ or a union of two $K3$s, and the twistor fiber $%
\mathbb{P}^{1}$ intersects the branch locus at eight points, and thus this
double cover is a fibration by a genus three Riemann surface which is the
double cover of the $\mathbb{P}^{1}$ fiber branched over eight points.
In \cite{Heckman:2013sfa} the branch locus of the double cover is a $K3$ and the twistor fiber
$\mathbb{P}^{1}$ intersects the branch locus at four points, thus that double
cover is a fibration by a genus one Riemann surface which is the double
cover of the $\mathbb{P}^{1}$ fiber branched over four points.\footnote{In this way a $T^2$ fiber is produced and this can be connected to F-theory configurations via general heterotic/F-theory dualities for example \cite{Morrison:1996na, Lerche:1999de, Friedman:1997yq, LopesCardoso:1996hq, Grimm:2012yq, Heckman:2013sfa}.} The
noncompact Calabi-Yau can be produced by deleting a $K3$ from the double cover
of the twistor space branched over $K3$. In this paper, compact Calabi-Yau are
produced by the double cover branched over the divisor of a different divisor
class.

\section{$K3$ fibered Calabi-Yau threefolds}

\vspace{1pt}\renewcommand{\theequation}{4.\arabic{equation}} %
\setcounter{equation}{0}

\label{sec_construction Kahler CY}\vspace{1pt}

\subsection{The $\bP^3=\tw(S^4)$ example}

In this subsection we shall construct a $K3$ fibered Calabi-Yau threefold $M$ with a rank $5$ stable bundle $V$ satisfying the following constraints

\begin{enumerate}
\item $\wedge^5 V \cong \sO_M$;
\item $c_2(V)=c_2(M)$;
\item $c_3(V)=6$.
\end{enumerate}

To fix the notation we first recall the construction of the $K3$ fibered Calabi-Yau threefold $M$. Let $X_1$ and $X_2$ be two smooth quartic $K3$ surfaces in $\bP^3$ so that their intersection $C=X_1\cap X_2$ is smooth. We shall specify the choice of these $K3$ surfaces later when we construct the stable bundle $V$. For now, we work with any such $K3$ surfaces. Obviously, $X_1$ and $X_2$ generates a pencil in the complete linear system $|\sO_{\bP^3}(4)|$ with fixed locus $C$. Let $\pi:\tilde\bP^3\to \bP^3$ be the blow-up of $\bP^3$ along $C$ with exception divisor $E$. By blowing up the fixed locus $C$, we obtain a $K3$-fibration $q:\tilde\bP^3\to \bP^1$. Because the normal bundle of $C\subset \bP^3$ is $\sO_C(4)\oplus \sO_C(4)$, we have an isomorphism $E\cong C\times \bP^1$. We denote by $\tilde X_i$ the proper transform of $X_i$, and $C_i=\tilde X_i\cap E$. See Figure 1.

\begin{figure}[ht!]

\centering
\begin{tikzpicture}[yscale=0.6,xscale=0.6]

\draw [black]  (2,0) -- (2,2);
\draw [black]  (2,0) -- (4,2);
\draw [black]  (2,2) -- (4,4);
\draw [black]  (4,2) -- (4,4);
\node [below] at (2.5,1.5) {\tiny $X_1$};

\draw [black]  (6,0) -- (6,2);
\draw [black]  (6,0) -- (4,2);
\draw [black]  (6,2) -- (4,4);
\draw [black]  (4,2) -- (4,4);
\node [below] at (5.5,1.5) {\tiny $X_2$};
\node [below] at (4,4.6) {\tiny $C$};

\draw [black]  (-8,0) -- (-7,1);
\draw [black]  (-8,0) -- (-8,3);
\draw [black]  (-8,3) -- (-7,4);
\draw [black]  (-7,1) -- (-7,4);
\node [below] at (-7.5,1.5) {\tiny $\tilde X_1$};
\node [below] at (-7,4.8) {\tiny $\tilde C_1$};

\draw [black]  (-7,4) -- (-3,4);
\draw [black]  (-8,3) -- (-4,3);
\node [below] at (-5.5,3.8) {\tiny $E$};

\draw [black]  (-4,0) -- (-3,1);
\draw [black]  (-4,0) -- (-4,3);
\draw [black]  (-4,3) -- (-3,4);
\draw [black]  (-3,1) -- (-3,4);
\node [below] at (-3.5,1.5) {\tiny $\tilde X_2$};
\node [below] at (-3,4.8) {\tiny $\tilde C_2$};
\node [left] at (-9,2) {$\tilde \bP^3$};
\node [right] at (7,2) {$\bP^3$};

\draw [black,help lines, ->] (-1,2) -- (1,2);
\node [below] at (0,2.6) {\tiny $\pi$};
\draw [black,help lines, ->] (-5.5,0) -- (-5.5,-1);
\node [right] at (-5.5,-0.5) {\tiny $q$};
\draw [black]  (-8,-1.5) -- (-3,-1.5);
\node [right] at (-7.5,-1.5) {\tiny $\bullet$};\node [above] at (-7.5,-1.5) {\tiny $Q_1$};
\node [left] at (-3.2,-1.5) {\tiny $\bullet$};\node [above] at (-3.2,-1.5) {\tiny $Q_2$};
\node [left] at (-9,-1.5) {$\bP^1$};
\end{tikzpicture}
          \label{fig:1}\caption{Blowup of $\bP^3$.}
\end{figure}

Let $f:M\to\tilde\bP^3$ be the double cover of $\tilde\bP^3$ branched along $\tilde X_1\cup\tilde X_2$. Then it is straightforward to verify that $M$ is a $K3$ fibered Calabi-Yau threefold. We denote the inverse image of $\tilde X_i$ by $\tilde X_i'$. See Figure 2.

\begin{figure}[ht!]

\centering
\begin{tikzpicture}[yscale=0.6,xscale=0.6]

\draw [black]  (2,0) -- (3,1);
\draw [black]  (2,0) -- (2,3);
\draw [black]  (2,3) -- (3,4);
\draw [black]  (3,1) -- (3,4);
\node [below] at (2.5,1.5) {\tiny $\tilde X_1$};
\node [below] at (3,4.8) {\tiny $\tilde C_1$};

\draw [black]  (3,4) -- (7,4);
\draw [black]  (2,3) -- (6,3);
\node [below] at (4.5,3.8) {\tiny $E$};

\draw [black]  (6,0) -- (7,1);
\draw [black]  (6,0) -- (6,3);
\draw [black]  (6,3) -- (7,4);
\draw [black]  (7,1) -- (7,4);
\node [below] at (6.5,1.5) {\tiny $\tilde X_2$};
\node [below] at (7,4.8) {\tiny $\tilde C_2$};
\node [right] at (8,2) {$\tilde \bP^3$};

\draw [black,help lines, ->] (5.5,0) -- (5.5,-1);
\node [right] at (5.5,-0.5) {\tiny $q$};
\draw [black]  (2,-1.5) -- (7,-1.5);
\node [left] at (3,-1.5) {\tiny $\bullet$};\node [above] at (3,-1.5) {\tiny $Q_1$};
\node [left] at (7,-1.5) {\tiny $\bullet$};\node [above] at (7,-1.5) {\tiny $Q_2$};
\node [right] at (8,-1.5) {$\bP^1$};

\draw [black]  (-8,0) -- (-7,1);
\draw [black]  (-8,0) -- (-8,3);
\draw [black]  (-8,3) -- (-7,4);
\draw [black]  (-7,1) -- (-7,4);
\node [below] at (-7.5,1.5) {\tiny $F_1$};
\node [below] at (-7,4.8) {\tiny $C_1$};

\draw [black]  (-7,4) -- (-3,4);
\draw [black]  (-8,3) -- (-4,3);
\node [below] at (-5.5,3.8) {\tiny $E$};

\draw [black]  (-4,0) -- (-3,1);
\draw [black]  (-4,0) -- (-4,3);
\draw [black]  (-4,3) -- (-3,4);
\draw [black]  (-3,1) -- (-3,4);
\node [below] at (-3.5,1.5) {\tiny $F_2$};
\node [below] at (-3,4.8) {\tiny $C_2$};
\node [left] at (-9,2) {$M$};

\draw [black,help lines, ->] (-1,2) -- (1,2);
\node [below] at (0,2.6) {\tiny $f$};
\draw [black,help lines, ->] (-5.5,0) -- (-5.5,-1);
\node [right] at (-5.5,-0.5) {\tiny $q$};
\draw [black]  (-8,-1.5) -- (-3,-1.5);
\node [right] at (-7.5,-1.5) {\tiny $\bullet$};\node [above] at (-7.5,-1.5) {\tiny $P_1$};
\node [left] at (-3.2,-1.5) {\tiny $\bullet$};\node [above] at (-3.2,-1.5) {\tiny $P_2$};
\node [left] at (-9,-1.5) {$\bP^1$};
\end{tikzpicture}
          \label{fig:2}\caption{Double cover of $\tilde\bP^3$.}
\end{figure}

\begin{figure}

\end{figure}


\begin{figure}

\end{figure}

\vskip 5pt

Having this Calabi-Yau threefold $M$, we compute its second Chern class $c_2(M)$. Let $\ell$ be the class of a line on $\bP^3$. Then $c_2(T_{\bP^3})=6\ell$. Consider the blowing-up $\pi:\tilde\bP^3\to\bP^3$. Let $\alpha\in H_2(\tilde\bP^3,\bZ)$ be the class of proper transform of $C$. Then $\alpha=[C_1]=[C_2]$. Consider the short exact sequence
\begin{equation}
0\map\pi^*\Omega_{\bP^3}\map\Omega_{\tilde\bP^3}\map\Omega_{\tilde\bP^3/\bP^3}\map 0.
\end{equation}
We obtain
\begin{equation}\label{chern}
c_2(\Omega_{\tilde\bP^3})=c_2(\pi^*\Omega_{\bP^3})+c_1(\pi^*\Omega_{\bP^3})\cdot c_1(\Omega_{\tilde\bP^3/\bP^3})+c_2(\Omega_{\tilde\bP^3/\bP^3}).
\end{equation}

Before we compute the expression of $c_2(\Omega_{\tilde\bP^3})$, we can get a rough picture on what it looks like by the following argument. Because $\Omega_{\tilde\bP^3/\bP^3}$ is a sheaf supported on $E$, its Chern classes can be localized on $E$. On the other hand, we have the following intersection numbers in cohomology of $\tilde\bP^3$
\begin{enumerate}
\item $\pi^*\ell\cdot [\tilde X_1]=4$;
\item $\alpha\cdot [\tilde X_1]=0$.
\end{enumerate}
Now we compute the intersection numbers of $[\tilde X_1]$ with both sides of (\ref{chern}). Noticing that
\begin{equation}
c_2(\Omega_{\tilde\bP^3})\cdot \tilde X_1=c_2(\Omega_{\tilde\bP^3}|_{\tilde X_1})=c_2(\Omega_{\tilde X_1})=24,
\end{equation}
and
\begin{equation}
c_2(\pi^*\Omega_{\bP^3})\cdot \tilde X_1 = \pi^*(6\ell)\cdot [\tilde X_1]=24,
\end{equation}
we know that $c_1(\pi^*\Omega_{\bP^3})\cdot c_1(\Omega_{\tilde\bP^3/\bP^3})+c_2(\Omega_{\tilde\bP^3/\bP^3})$ is a class supported on $E$ and has zero intersection with $\tilde X_1$, which implies that it is equal to $\rho \alpha$ for some integer $\rho$.

Now we compute this number $\rho$. Recall that $E$ is the exceptional divisor of $\pi:\tilde\bP^3\to \bP^3$ and $E\cong C\times \bP^1$. Let $\iota: E\to \tilde\bP^3$ be the natural immersion and $q_1:E\to C$ and $q_2:E\to \bP^1$ be the projections. Then we have
\begin{equation}\label{relOmega}
\Omega_{\tilde\bP^3/\bP^3}=\iota_*\Omega_{E/C}=\iota_*q_2^*\sO_{\bP^1}(-2)=\iota_*\sO_E(-2C_1).
\end{equation}

From the short exact sequence
\begin{equation}
0\map\sO_{\tilde\bP^3}(-E)\map\sO_{\tilde\bP^3}\map\iota_*\sO_E\map 0,
\end{equation}
we obtain
\begin{displaymath}
\begin{matrix}
\ch(\iota_*\sO_E)&=&\ch(\sO_{\tilde\bP^3})-\ch(\sO_{\tilde\bP^3}(-E))\\
&=&0+E-\frac{1}{2}E^2+\frac{1}{6}E^3.
\end{matrix}
\end{displaymath}
It follows from (\ref{relOmega}) that
\begin{displaymath}
\begin{array}{rcl}
\ch(\Omega_{\tilde\bP^3/\bP^3})&=&\ch(\iota_*\sO_E(-2C_1))\\
&=&\ch(\iota_*\sO_E)\ch(\sO_{\tilde\bP^3}(-2\tilde X_1))\\
&=&(0+E-\frac{1}{2}E^2+\frac{1}{6}E^3)(1-2\tilde X_1+\frac{1}{2}(-2\tilde X_1)^2+\frac{1}{6}(-2\tilde X_1)^3)\\
&=&0+E-\frac{1}{2}E^2-2E\tilde X_1+\frac{1}{6}E^3+E^2\tilde X_1+2E\tilde X_1^2.
\end{array}
\end{displaymath}
Therefore
\begin{equation}
c_1(\Omega_{\tilde\bP^3/\bP^3})=E,\quad\and\quad c_2(\Omega_{\tilde\bP^3/\bP^3})=E^2+2E\tilde X_1
\end{equation}
Noticing that $c_1(\pi^*\Omega_{\bP^3})=-\tilde X_1-E$, by (\ref{chern}) we obtain
\begin{equation}
c_2(\Omega_{\tilde\bP^3})=c_2(\pi^*\Omega_{\bP^3})+(-\tilde X_1-E)E+E^2+2E\tilde X_1=\pi^*(6\ell)+E\tilde X_1.
\end{equation}
Since $E\tilde X_1=[C_1]=\alpha$, we get $\rho=1$ and
\begin{equation}
c_2(\Omega_{\tilde\bP^3})=6\pi^*\ell+\alpha.
\end{equation}

To compute $c_2(M)$, we consider the double cover $f: M\to\tilde\bP^3$. Using the short exact sequence
\begin{equation}
0\map f^*\Omega_{\tilde\bP^3}\map\Omega_M\map\Omega_{M/\tilde\bP^3}\map 0,
\end{equation}
we have
\begin{equation}\label{chern2}
c_2(\Omega_{M})=c_2(f^*\Omega_{\tilde\bP^3})+c_1(f^*\Omega_{\tilde\bP^3})\cdot c_1(\Omega_{M/\tilde\bP^3})+c_2(\Omega_{M/\tilde\bP^3}).
\end{equation}
Because $\Omega_{M/\tilde\bP^3}$ supports at $\tilde X_1'\cup \tilde X_2'$, simple computation shows that
\begin{equation}
c_1(\Omega_{M/\tilde\bP^3})=2[\tilde X_1'],\quad c_2(\Omega_{M/\tilde\bP^3})=0.
\end{equation}
It follows that $c_1(f^*\Omega_{\tilde\bP^3})\cdot c_1(\Omega_{M/\tilde\bP^3})=0$. Hence
\begin{equation}
c_2(\Omega_{M})=c_2(f^*\Omega_{\tilde\bP^3}).
\end{equation}
In conclusion, we obtain
\begin{equation}
c_2(M)=c_2(\Omega_M)=c_2(f^*\Omega_{\tilde\bP^3})=6f^*\pi^*\ell+f^*\alpha.
\end{equation}

Next we construct a rank $5$ bundle $V$ over $M$ satisfying the conditions list at the beginning of this section. We shall construct $V$ as a direct sum
\begin{equation}
V=V_2\oplus V_3
\end{equation}
so that $V_2$ is a rank $2$ stable bundle with
\begin{enumerate}
\item $\wedge^2 V_2 \cong \sO_M$;
\item $c_2(V_2)=6f^*\pi^*\ell$;
\item $c_3(V_2)=0$,
\end{enumerate}
and $V_3$ is a rank $3$ stable bundle with
\begin{enumerate}
\item $\wedge^3 V_3 \cong \sO_M$;
\item $c_2(V_3)=f^*\alpha$;
\item $c_3(V_3)=6$.
\end{enumerate}
Once can verify easily that any such $V$ satisfies the required conditions.

To construct $V_2$, we recall that for any $d>0$, one can find a rank $2$ instanton bundle $W_d$ over $\bP^3$ which are stable with
\begin{equation}
\wedge^2 W_d=\sO_{\bP^3},\quad c_2(W_d)=d\ell,\quad \and\quad c_3(W_d)=0.
\end{equation}
In our case, we take $V_2=f^*\pi^*W_6$.

For $V_3$, we shall use the construction of stable bundles on Calabi-Yau threefolds in \cite{Wu:2014lka}. For convenience, we state the main theorem in \cite{Wu:2014lka} for the special case of rank $3$ bundles over $K3$ fibered Calabi-Yau threefolds as follows
\begin{thm}\label{wy}\cite{Wu:2014lka}
Let $M\to \bP^1$ be a $K3$-fibered Calabi-Yau threefold. Let $\{Y_i\}$ be disjoint irreducible curves in distinct fibers of $M$. Suppose $g(Y_i)\ge 1$. Then there exists a rank $3$ stable bundle $W$ over $M$ with
\begin{enumerate}
\item $\wedge^3 W \cong \sO_M$;
\item $c_2(W)=\sum[Y_i]$;
\item $c_3(W)=\sum(2g(Y_i)-2)$.
\end{enumerate}
\end{thm}

To apply this theorem to our case, we need to choose $K3$ surfaces $X_1$ and $X_2$ in $\bP^3$ carefully and find curves $Y_i$ satisfying conditions
\begin{enumerate}
\item $\sum[Y_i]=f^*\alpha$;
\item $\sum(2g(Y_i)-2)=6$.
\end{enumerate}
To achieve this, we use the following theorem of Mori \cite{Mori}
\begin{thm}\label{mori:curve}\cite{Mori}
There exists a non-singular curve of degree $d>0$ and genus $g\ge 0$ on a non-singular quartic surface in $\bP^3$ if and only if (1) $g=\frac{d^2}{8}+1$ or (2) $g<\frac{d^2}{8}$ and $(d,g)\ne(5,3)$.
\end{thm}
Now we let $d=4$ and $g=1$. They satisfy condition (2) in the above theorem. So we can find a smooth quartic $K3$ surface $X_1$ and a smooth curve $C_1'\subset X_1$ with $\deg C_1'=4$ and $g(C_1')=1$. Similarly, taking $d=8$ and $g=4$, we get another smooth quartic $K3$ surface $X_2$ and curve $C_2'\subset X_2$ with $\deg C_2'=8$ and $g(C_2')=4$. We can also choose such $X_1$ and $X_2$ so that they intersect along a smooth curve $C$.

Since $H_2(\bP^3,\bZ)=\bZ$ and $C$ has degree $16$, we know that $[C]=[4C_1']=2[C_2']$. By theorem \ref{wy}, we can find a rank $3$ stable bundle $V_3$ that fulfills the requirement.

\subsection{Physical interpretations}

In this subsection we make some physical interpretations of the model after
compactification to four dimensions. The commutant of this rank 5
bundle in $E_{8}$ is the $SU(5)$ grand unified group. Thus this gives rise
to a supersymmetric $SU(5)$ GUT model with three generations of chiral fermions.

In type IIB and F-theory, the GUT symmetry breaking can be obtained by
an internal gauge field flux. In these duality frames, the gauge theory degrees of freedom can be
packaged onto the worldvolume of seven-branes wrapping a divisor inside the
base of the elliptic Calabi-Yau four-folds. There are examples in which the
base is $\mathbb{P}^{3}$ and the divisor is $\mathbb{P}^{1}\times \mathbb{P}%
^{1}$ \cite{Beasley:2008kw, Donagi:2008kj, Buican:2006sn, Klemm:1996ts}.
There is only one generator $H$ for the second homology of $\mathbb{P}^{3}$ while there are two generators $\sigma_{1}$
and $\sigma_{2}$ for the two $\mathbb{P}^{1}$s of $\mathbb{P}^{1}\times \mathbb{P}^{1}$. This means that
the two-cycle $\sigma =\sigma _{1}-\sigma _{2}$ trivializes in $\mathbb{P}%
^{3}$. We can turn on an abelian internal gauge field flux along the linear combination of the
two $\mathbb{P}^{1}$s of the divisor $\mathbb{P}^{1}\times \mathbb{P}^{1}$ in $%
\mathbb{P}^{3}$, which breaks the GUT symmetry group to Standard Model
group, see for example \cite{Buican:2006sn, Beasley:2008kw, Donagi:2008kj}.

Now we consider the scenario under the blow up and double cover map. The preimage of the two-cycle $(\pi{\circ}f)^{\ast
}(\sigma _{1}-\sigma _{2})$ trivializes in $M$. We can turn on the internal gauge field flux
along this non-homological two-cycle to break the symmetry to the
Standard Model group. This is reminiscent to the scenarios in for example \cite{Buican:2006sn, Beasley:2008kw, Donagi:2008kj}.

The model-dependent axions come from the $B$-field along the homologically
non-trivial two-cycles, while the internal gauge field flux is along the homologically
trivial two-cycle, hence their topological coupling vanishes after
integration on the internal manifold $M$. Due to this topological
mechanism, all couplings to the bulk axions automatically vanish. This
includes both the universal axion of the heterotic compactification and the model-dependent axions from the harmonic two-forms of $M$.

The $c_{3}=6, g=4$ case is the example of three generation models with rank
5 bundles. This corresponds to supersymmetric $SU(5)$ GUT models with three
generations of chiral fermions in four dimensions. After a GUT symmetry
breaking, some of the models can give rise to models with Standard Model
gauge groups and three generations of chiral fermions.

We have also constructed rank 5 bundles $V$ with $c_{3}=12$. For the $%
\mathbb{P}^{3}$ case, it is possible to have freely acting $\mathbb{Z}_{2}$
involution, so that the $c_{3}=12, g=7$ case may also be an example of a
three generation model, after $\mathbb{Z}_{2}$ involution. The
classification of the automorphism groups of the $K3$-fibered Calabi-Yau here
is an interesting future direction.

This construction can potentially be generalized for other twistor spaces Tw$(n\mathbb{P}%
^{2})$, with $n=1,2,3$. In this way, the double cover of the blow up of Tw$(n\mathbb{P}%
^{2})$, with $n=0,1,2,3$, are the $K3$ fibrations over $\mathbb{P}^{1}$. We can
construct the rank 2 stable instanton bundle $V_{2}$ on the $K3$ fibration over
$\mathbb{P}^{1}$ in similar ways. Once we identify the appropriate curves in
the $K3$ fibers, we can construct the rank 3 stable bundle $V_{3}$ in an
analogous way \cite{Wu:2014lka}. The details of the construction of $V_{3}$ for general $n$ is
beyond the scope of the present paper and is an interesting direction for
future investigation.

\section{Non-K\"{a}hler Calabi-Yau spaces}

\vspace{1pt}\renewcommand{\theequation}{5.\arabic{equation}} %
\setcounter{equation}{0}

\label{sec_construction non-Kahker}\vspace{1pt}

\subsection{Double cover of $\tw(2\bP^{2})$}

\label{subsec_2P2 detailed}\vspace{1pt}

It is known that \cite{Poon86} the twistor space $\tw(2\bP^{2})$ is a crepant resolution of the intersection of two quadrics in $\bP^{5}$. Explicitly, we let
\begin{equation}
\begin{array}{l}
Q=\{z\in \bP^{5}:%
\,z_{0}^{2}+z_{1}^{2}+z_{2}^{2}+z_{3}^{2}+z_{4}^{2}+z_{5}^{2}=0\} \\
Q_{\lambda}=\{z\in \bP^{5}:\,2z_{0}^{2}+2z_{1}^{2}+\lambda z_{2}^{2}+\frac{3}{2%
}z_{3}^{2}+z_{4}^{2}+z_{5}^{2}=0\}%
\end{array}%
\end{equation}%
for real number $\frac{3}{2}<\lambda <2$. Let $Z_0=Q\cap Q_{\lambda}$. Then $Z_0$ is a projective threefold with $4$ ordinary double point singularities. These $4$ double points are
\begin{equation}
(0,0,0,0,1,\sqrt{-1}),(0,0,0,0,1,-\sqrt{-1}),(1,\sqrt{-1},0,0,0,0),(1,-\sqrt{%
-1},0,0,0,0).  \notag
\end{equation}%
The twistor space $Z=\tw(2\bP^{2})$ is a small resolution $Z\to Z_0$ of $Z_{0}$ at these $4$ ordinary double points.

To understand better the geometry of the double cover Calabi-Yau threefold, we consider the following general phenomenon. Let $D\subset X$ be a very ample divisor of a smooth variety $X$. Let $S$ be a general member in the linear system $|\sO_X(2D)|$. Consider the pencil generated by the nonreduced divisor $2D$ and $S$. We denote the tautological family by $Y'\to X\times\bP^1$ so that $Y'_0=2D$ and $Y'_1=S$. Applying semistable reduction, we can replace the nonreduced central fiber $Y'_0$ by a reduced one. Precisely, we can find a base change $b:T\to\bP^1$ and take the normalization of the
pullback family, so that the resulting family $Y\to X\times T$ satisfies $Y_0$ is reduced. In fact $Y_0$ is a double cover of $D$.

To see this, we consider the degeneration of $X$ as follows. Let $\tilde {X}\to X\times\bP^1$ be the blow-up of $X\times\bP^1$ along $D\times 0$. Then as a family parameterized by $\bP^1$, $\tilde {X}_t\cong X$ for $t\ne 0$ and $\tilde {X}_0$ is the union $X\cup_D P(\mathcal{O}_D\oplus N_D)$, where $N_D$ is the normal bundle of $D$. Since $S\subset X$ is normal (or transverse) to $D$, that is, the natural homomorphism $\mathcal{I}_S\otimes_{\sO_X} \sO_D\to\sO_D$ is injective, applying the argument in \cite{LW11}, we can modify the family $Y\to X\times\bP^1$ to obtain $Y\subset \tilde {X}$, possibly after a base change, so that $Y_0\subset P(\sO_D\oplus N_D)$ and it is normal to $D$. Hence, it is a double cover of $D$. See Figure 3.

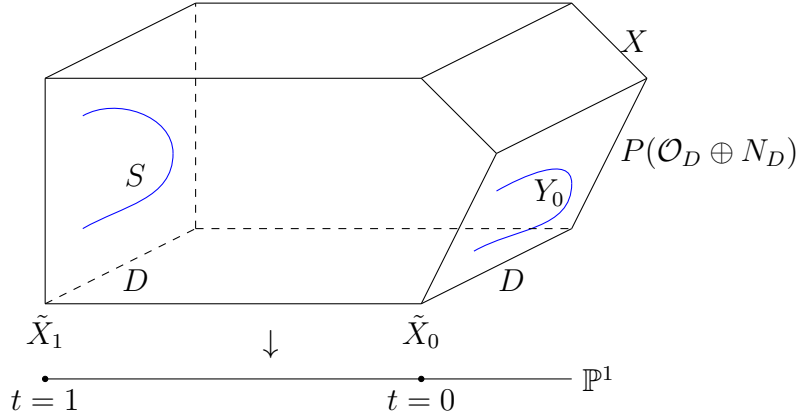
\begin{figure}[ht!]
\centering
\begin{tikzpicture}[yscale=1,xscale=1]

\draw [black]  (0,0) -- (0,3);
\draw [dashed,black]  (0,0) -- (2,1);
\draw [black]  (0,3) -- (2,4);
\draw [dashed, black]  (2,1) -- (2,4);

\draw [blue]  (.5,1) to [out=30,in=270] (1.7,2);
\draw [blue]  (.5,2.5) to [out=30,in=90] (1.7,2);

\draw [blue]  (5.7,0.7) to [out=30,in=270] (7,1.6);
\draw [blue]  (6,1.5) to [out=30,in=90] (7,1.6);

\draw [black]  (5,0) -- (6,2);
\draw [black]  (6,2) -- (5,3);
\draw [black]  (5,0) -- (7,1);
\draw [black]  (5,3) -- (7,4);
\draw [black]  (7,1) -- (8,3);
\draw [black]  (8,3) -- (7,4);
\draw [black]  (6,2) -- (8,3);

\draw [black]  (0,0) -- (5,0);
\draw [black]  (0,3) -- (5,3);
\draw [black]  (2,4) -- (7,4);
\draw [dashed, black]  (2,1) -- (7,1);

\node [below] at (3,-.2) {$\downarrow$};
\draw [black]  (0,-1) -- (7,-1);

\node [right] at (7,-1) {$\bP^1$};
\node [below] at (5,-1) {$t=0$};
\filldraw (0,-1) circle [radius=1pt];
\filldraw (5,-1) circle [radius=1pt];
\node [below] at (0,-1) {$t=1$};
\node [below] at (0,0) {$\tX_1$};
\node [below] at (5,0) {$\tX_0$};
\node [right] at (7.5,3.5) {$X$};
\node [right] at (7.5,2) {$P(\sO_D\oplus N_D)$};
\node [below] at (1.2,.6) {$D$};
\node [below] at (6.2,.6) {$D$};
\node [below] at (1.2,2) {$S$};
\node [below] at (6.7,1.8) {$Y_0$};
\end{tikzpicture}
\caption{Double cover.}
\label{fig:3}
\end{figure}

To apply this construction to the twistor space case, we let $X=Q_{\lambda}$, $D=Q\cap Q_{\lambda}$ and $S=S_4\cap Q_{\lambda}$ for a general quartic hypersuface $S_4\subset \bP^5$. Then we can find a family $Y$ of Calabi-Yau threefolds parameterized by $\bP^1$, so that the fiber $Y_1=S$ is a type $(2,4)$ Calabi-Yau threefold in $\bP^5$, and $Y_0$ is a Calabi-Yau threefold which is a double cover of $D$. A small resolution $M\to Y_0$ gives a smooth Calabi-Yau threefold $M$.

It follows that $M$ and $Y_{t}$ are related by a conifold transition. Here $M$ is a non-K\"{a}hler
Calabi-Yau, while $Y_{t}$ is a K\"{a}hler Calabi-Yau.

We obtain the following:
\begin{prop}
\label{rela} There exists a family of Calabi-Yau threefolds $Y$ parameterized by $\bP^1$, so that $Y_t$ is a type $(2,4)$ smooth projective Calabi-Yau threefold in $\bP^5$, and $Y_0 $ is a singular Calabi-Yau threefold with $8$ ordinary double points. The resolution of these double points gives a Moishezon (non-K\"{a}hler) Calabi-Yau threefold which is a double cover of the twistor space $\tw(2\bP^2)$.
\end{prop}

\subsection{Vector bundles over non-K\"{a}hler Calabi-Yau}

In this subsection we consider certain rank $4$ bundles over the double cover Calabi-Yau threefold constructed above.

We first discuss the tangent bundle of the double cover $M$ and its Chern classes. The intersection $Z_{0}=Q\cap Q_{\lambda}$ has $4$ ordinary double points, taking double covering, the resulting $Y_{0}$ has $8$ ordinary double points. Let $M\to Y_0$ be a crepant resolution of these $8$ points with exceptional $\bP^1$s.

Let $H=O_{\bP^{5}}(1)$. The Chern class of type $(2,4)$ Calabi-Yau threefold family $Y_{t} \subset \bP^{5}$ follows from expanding to third order in the divisor class,%
\begin{equation}
c(Y_{t})=\frac{(1+H)^{6}}{(1+2H)(1+4H)}=1+7H^{2}-22H^{3},
\end{equation}%
The Chern invariants are%
\begin{equation}
c_{1}(Y_{t})=0\text{, \ \ }c_{2}(Y_{t})=7H^{2}\text{, \
\ }c_{3}(Y_{t})=-22H^{3}.
\end{equation}%
The integrated Chern class is
\begin{equation}
\chi (Y_{t})=\int \text{\ }c_{3}(Y_{t})=-176\text{,}
\end{equation}%
since $H^3|_{Y_t}=8.$

We denote $p$: $M\to \bP^{5}$ the natural map and $H_{M}=p^{\ast }(H_{\bP^{5}})$. The Chern classes of $M$ is
therefore
\begin{equation}
c_{1}(M)=0,~~~c_{2}(M)=7H_{M}^{2}+%
\sum_{i=1}^{8}E_{i},~~~c_{3}(M)=-22H_{M}^{3}+8\chi (\bP%
^{1})=-22H_{M}^{3}+16,
\end{equation}%
where $E_{i}$s are the Poincare duals of the 8 $\bP^{1}$s. The
integrated Chern class is%
\begin{equation}
\chi (M)=\int \text{\ }c_{3}(M)=-160\text{.}
\end{equation}

Next we consider vector bundles on $M$. We can construct various vector bundles with either nonzero $c_{3}$ or zero $c_{3}$. The examples with nonzero $c_{3}(V)$ include the tangent bundle $T_{M}$, while the examples with zero $c_{3}(V)$ include the instanton bundles.

We see that $c_{2}(M)-c_{2}(V)$ is the total class of the minimal effective
curves $[W]=\sum_{i=1}^{8}E_{i}$. The $c_{2}(M)-c_{2}(V)$ is the class of
the minimal effective curves on $M$,%
\begin{equation}
c_{2}(M)-c_{2}(V_{M})=[W].
\end{equation}%
The fivebranes can wrap on these effective curves, which are the 8 $\mathbb{P%
}^{1}$s in $M$.

There exists stable instanton bundles with rank $2l$ on $\bP^{2l+1}$ \cite{AO}. We are considering the $l=2$ case, that is, the rank $4$ instanton bundles $V_m$ of quantum number $m\in \bZ$ on $\bP^{5}$. The total Chern class of this vector bundle is
\begin{equation}
c(V_m)=(1-H^{2})^{-m}.
\end{equation}%
We define $V=p^{\ast }(V_m)$. The Chern class for the vector bundle of the three-fold $M$ are expanded to third
order in the divisor class,
\begin{equation}
c(V)=(1-H_{M}^{2})^{-m}=1+mH_{M}^{2}.
\end{equation}%
The Chern invariants are%
\begin{equation}
c_{1}(V)=0\text{, \ \ }c_{2}(V)=mH_{M}^{2}\text{, \ \ }%
c_{3}(V)=0.
\end{equation}%
Hence we set $m=7$, and we see that $c_{2}(M)-c_{2}(V)=[W]$ for a total minimal effective class $[W]$.

By using the theorems in \cite{AO}, we expect the bundle $V$ is stable. It suffices to show that $H^{0}(M,(\wedge^{q}V)_{\mathrm{norm}})=0$ for $0<q<4$. Recall that \cite{AO} the instanton bundle $V_m$ on $\bP^5$ satisfies the exact sequences including
\begin{eqnarray}
0 &\map &\sO(-1)^{m}\map S^{\vee }\map
V_m\map 0, \\
0 &\map &\sO(-1)^{m}\map \sO%
^{2l+2m}\map S\map 0,
\end{eqnarray}%
where $V_m$ is the vector bundle with rank $2l$, and $S^{\vee }$ is a Schwarzenberger bundle of rank $2l+m.$ We are in the situation with $l=2$ and $m=7.$ So we have that%
\begin{eqnarray}
0 &\map &\sO(-1)^{7}\map S^{\vee }\map
V_7\map 0, \\
0 &\map &\sO(-1)^{7}\map \sO^{18}\map
S\map 0.
\end{eqnarray}%
Therefore $S$ is rank 11, and $V_7$ is rank 4. We also have the exact
sequences,
\begin{eqnarray}
0 &\map &\mathrm{Sym}^{2}(\sO(-1)^{7})\map \sO%
(-1)^{7}\otimes S^{\vee }\map \wedge ^{2}S^{\vee }\map \wedge
^{2}V_7\map 0, \\
0 &\map &\wedge ^{2}S^{\vee }\map \wedge ^{2}\sO%
^{18}\map \sO^{18}\otimes \sO(1)^{7}\map
\mathrm{Sym}^{2}(\sO(1)^{7})\map 0,
\end{eqnarray}%
which can be used to compute $H^{0}(M,(\wedge^{q}V)_{\mathrm{norm}})$. In the end, using the method of \cite{AO}, one can show the stability of $V$ in this way.


\section{Discussion}

\renewcommand{\theequation}{9.\arabic{equation}} \setcounter{equation}{0}

\label{sec_discussion}

\vspace{1pt}

In this paper we constructed K\"{a}hler and non-K\"{a}hler Calabi-Yau
manifolds and propose to use them in the heterotic string compactifications.
We constructed these manifolds from branched double covers of twistor spaces
of four-manifolds with self-dual conformal structures. These manifolds as the
branched covers solve the conformally balanced equation. We also constructed
stable and polystable vector bundles on these manifolds, which satisfy the
anomaly cancellation condition and the Hermitian-Yang-Mills equations. Some
classes of the vector bundles that we constructed give rise to supersymmetric grand unified models
with three generations of chiral fermions.

Our construction includes both K\"{a}hler Calabi-Yau spaces and non-K\"{a}hler
Calabi-Yau spaces. We constructed them by branched double covers of twistor spaces, or branched
double covers of the blow-ups of the twistor spaces. In the latter case, these
Calabi-Yau manifolds are also $K3$ fibrations over $\mathbb{P}^{1}$. We
considered the twistor spaces of the connected sum of $n$ $\mathbb{P}^{2}$s.
The twistor spaces \textrm{Tw}$(n\mathbb{P}^{2})$ with $n=0,1,$ which are $%
\mathbb{P}^{3}$ and flag manifold respectively, are K\"{a}hler, while the
\textrm{Tw}$(n\mathbb{P}^{2})$ with $n=2,3$ are non-K\"{a}hler. The branch
locus of the double cover of the twistor spaces \textrm{Tw}$(n\mathbb{P}^{2})$ can be either smooth or singular. In the smooth case, the
branch locus is a divisor whose divisor class is twice of the divisor class
of $K3$ surface. In the singular case, the branch locus is a union of two $K3$
surfaces, and after blowing up along the singular locus, the resulting manifold
is a $K3$ fibration over $\mathbb{P}^{1}$. In this way, we can uniformly treat
these twistor spaces \textrm{Tw}$(n\mathbb{P}^{2})$.

We have also constructed $K3$-fibered Calabi-Yau manifolds by the double covers of
the blow-ups of the twistor spaces. In the $K3$-fibered Calabi-Yau here, the $K3$
fiber may not be an elliptic $K3$, and hence the $K3$-fibered Calabi-Yau
here generally can be different from the elliptic Calabi-Yau with a Hirzebruch surface
base.

The compatification on these non-K\"{a}hler Calabi-Yau manifolds contains both
geometric moduli from the manifolds and the vector bundle moduli. Comparing
to the K\"{a}hler Calabi-Yau case, the non-K\"{a}hler Calabi-Yau compactifications in some cases may have potentially fewer
geometric moduli and bundle moduli than the K\"{a}hler case, and the
problem to stabilize these bundle moduli to particular values may
be an interesting direction. 

One of the interesting aspects of the heterotic compactification is to
consider the worldsheet instantons and the superpotential generated by them.
Comparing to K\"{a}hler Calabi-Yau case, the non-K\"{a}hler Calabi-Yau manifolds in some cases may have
potentially less rational curves. It hence may be interesting to consider the
worldsheet instantons in these non-K\"{a}hler Calabi-Yau manifolds.

Our approach to use the branched double covers of twistor spaces to produce K\"{a}hler and
non-K\"{a}hler Calabi-Yau manifolds and use them to compactify the heterotic
string theory to four dimensions opens new possibilities to construct
realistic Standard Model like physics from superstring theory. These
K\"{a}hler and non-K\"{a}hler Calabi-Yau manifolds lead to new compactifications that can give
rise to chiral matter with three generations and may be promising in phenomenologically viable
four-dimensional theory.

\section*{Acknowledgments}


\appendix

\renewcommand{\theequation}{A.\arabic{equation}} \setcounter{equation}{0}


This work was supported in part by NSF grant DMS-1159412, NSF grant
PHY-0937443, NSF grant DMS-0804454, and in part by the Fundamental Laws
Initiative of the Center for the Fundamental Laws of Nature, Harvard University. We would also like to thank T. Fei for discussion or correspondence.



\end{document}